\documentclass[%
12pt,T
 oneocolumn,
 nofootinbib,
 amsmath,amssymb,
 aps,
floatfix,
unsortedaddress
]{revtex4-2}

\usepackage{graphicx,color}
\usepackage{subfig}
\usepackage{dcolumn}
\usepackage{bm}
\usepackage{graphicx,color}
\usepackage{float}
\usepackage{romannum}
\usepackage{amsmath}
\usepackage{mathtools}
\usepackage{upgreek}
\usepackage{cancel}
\usepackage[pdfpagelabels, pdfencoding=auto, psdextra]{hyperref}
\hypersetup{%
 pdfsubject=Paper,
 pdfkeywords={nuclear physics} {Pionless EFT} {Two-Nucleon} {Lattice} {L\"usher formula},
 unicode = true,
 breaklinks = true,
 colorlinks = true,
 linkcolor = blue,
 citecolor = blue,
 menucolor = blue,
 citecolor = blue,
 urlcolor = blue
}

\graphicspath{{./}}

\begin{document}
	\pagenumbering{arabic}

\title{  Femtoscopic study of the $ \Omega \alpha $ interaction in heavy-ion collisions}
\author{Faisal Etminan}
 \email{fetminan@birjand.ac.ir}
\affiliation{
 Department of Physics, Faculty of Sciences, University of Birjand, Birjand 97175-615, Iran
}%
\affiliation{ Interdisciplinary Theoretical and Mathematical Sciences Program (iTHEMS), RIKEN, Wako 351-0198, Japan}

\date{\today}%

\begin{abstract}
 The two-particle momentum correlation  between the Omega-baryon ($\Omega$) and  the $^{4}He (\alpha) $ in high-energy heavy ion collisions  is explored.  
 Such correlations as an alternative source of information can help us further understand the interaction between $ \Omega $ and nucleons (N). 
 $ \Omega\alpha $ potentials in the single-folding potential approach are constructed by employing  two different  available $\Omega N $ interactions in $^{5}S_{2}$ channel, i.e, one is based on the (2 + 1)-flavor lattice QCD simulations near the physical point by the HAL QCD collaboration, and the other is based on the meson exchanges with effective Lagrangian, where in the latter case  the effect of coupled channels is considered. 
 It is found that the correlation functions at small size source depends on the potential model used. This implicitly means that at high density nuclear medium, $ \Omega \alpha $ momentum correlation could carry information on the feature of $ \Omega N $ interactions.
  Moreover, by extracting the scattering length and the effective range from obtained $ \Omega\alpha $ potentials,  the correlation functions are calculated within the Lednicky-Lyuboshits (LL) formalism.
  It is shown that since the $\Omega\alpha $ has large interaction range, 
  the LL  formula leads to different results at small source sizes.
 \end{abstract}


\maketitle
\section{Introduction} \label{sec:intro}
The spin-$ 2 $ Omega-nucleon ($ \Omega N $) state with $ S=-3 $ is expected to lack a repulsive core since 
the Pauli exclusion principle does not act between quarks in this channel~\cite{etminan2014}.
 The theoretical improvement of the HAL QCD Collaboration methods~\cite{Ishii2007,ishii2012,aoki2013,sasaki2020,etminan2024prd} coupled with  progress of high performance
 computing facilities provide the obtained hadron-hadron interactions at nearly physical quark masses from first principle lattice  QCD simulations~\cite{Gongyo2018,yanPrl2021,yan2022prd,yanPRL2023}. In the case of $ \Omega N $ system, they report a strongly attractive interaction 
 in the $ ^5 {S}_{2} $ channel~\cite{Iritani2019prb}.

Moreover, in Ref.~\cite{Sekihara2018} Sekihara, Kamiya and Hyodo have developed a model based on the meson exchanges (ME) with effective Lagrangians to investigate the origin of the attraction in the $ \Omega N $ interaction in the $ ^{5}S_{2} $ channel. They formulated an equivalent local potential for the $ \Omega N\:^{5}S_{2} $ interaction that reproduces $ \Omega N\:^{5}S_{2} $ scattering length $ 7.4\pm1.6 $ fm for the time range $ t/a=11 $ of the lattice simulations at nearly physical quark masses~\cite{doi2018,Iritani2019prb}  but with hadron masses tuned to the lattice simulations. The long range part of the potential is built on the exchanges of the $ \eta $ meson and correlated two mesons in the scalar-isoscalar channel (Known as $ "\sigma" $). The short-range part is constructed by the contact interaction. Furthermore, they considered the coupled-channel effects on $ \Omega N\:^{5}S_{2} $ interaction by adding the box diagrams with intermediate $ \Lambda\Xi,\Sigma\Xi $ and $ \Lambda\Xi \left(1530\right) $ channels. They concluded that even though the elimination of these channels induces the energy dependence of the single-channel $ \Omega N $ interaction, this effect is not significant.

High energy heavy ion collisions are an excellent method for creating heavy hadrons and  light (anti)nuclei, includes  molecular states made of various hadrons or compact system. 
One method for studying the hadron-hadron interaction that is hard to investigate in scattering experiment is measuring the 
momentum correlation functions in high-energy collisions~\cite{cho2017exotic}.
 It can provide information on both the effective emission source and the interaction potential.

The first measurement of the
proton-$\Omega$ correlation function \cite{Morita2016,mendez2024pomegafemtoscopy} in heavy-ion collisions by the STAR experiment  at the Relativistic Heavy-Ion Collider (RHIC)~\cite{STAR2019,alice2020unveiling} favors the proton-$\Omega$ bound state hypothesis. In Refs.~\cite{Fabbietti:2020bfg,alice2020unveiling} it is mentioned that the HAL QCD potential 
 is the potential most consistent with the LHC ALICE data. 

As the next step in the femtoscopic analyses, the hadron-deuteron correlation functions would be promising~\cite{PhysRevX.14.031051}. 
The production of $\Omega NN$ and $\Omega \Omega N$ in ultra-relativistic heavy-ion collisions using the Lattice QCD $\Omega N $, $\Omega \Omega$ potentials has been studied in Ref.~\cite{zhang2021production}.
And very recently, 	the momentum correlation between $ \Lambda \alpha $~\cite{jinno2024femtoscopic}  and $ \Xi \alpha $ ~\cite{kamiya2024} are examined to shed light on the interaction between a hyperon and nucleons $ \left(N\right) $.

Therefore, motivated by the above discussions, in this work, I want to explore the $ \Omega \alpha $ correlation function in the relativistic heavy ion collisions to probe the nature of $ \Omega N $ interactions as an independent source of information. 
The purpose of this work is to give an illustration for what can be expected from measuring $ \Omega \alpha $ correlations.
Since this is an  exploratory study, the techniques used are simple.

 A Woods-Saxon (WS) type form for $ \Omega +\alpha $ potential is obtained in the framework of single-folding potential (SFP) approach. Because the $ \alpha $-cluster is strongly bounded and not likely to change its properties, it is supposed that $\Omega $ and $\alpha$  move in an effective $\Omega \alpha$ potential.   
   The effective $ \Omega +\alpha $ nuclear potential is estimated by the single-folding of nucleon density in
 the $\alpha$-particle and $ \Omega N $ interaction~\cite{Satchler1979,10.1143/ptp.117.251,Miyamoto2018,Etminan:2019gds}. 
  Next, the obtained $\Omega \alpha$ potential is fitted to an analytical Woods-Saxon type function. 
  Then, the Schr\"{o}dinger equation is solved by the given $\Omega \alpha$ potential as the input to calculate binding energy and scattering phase shift. 
 And finally, the predictions on $\Omega \alpha$ momentum  correlation functions for given potentials are made by using the scattering wave functions of $\Omega \alpha$ system.
 
The paper is organized in the following way:
In Sec.~\ref{sec:folding-Model},  HAL QCD and meson-exchange $\Omega N$ potentials are introduced. 
Also the single-folding potential approach is described briefly. 
In Sec.~\ref{sec:Two-particle-CF} the formalism for two-particle momentum correlation
functions is concisely reviewed.  
Results and discussions for $ \Omega \alpha $ are presented in Sec.~\ref{sec:result}.
 The summary and  conclusions  are given in Sec.~\ref{sec:Summary-and-conclusions}.  
\section{$ \Omega N $ interactions  and single-folding potential approach } \label{sec:folding-Model}
In this section, the HAL QCD  and the meson-exchange $\Omega N$  potentials, which are used to find effective potential of the $ \Omega + \alpha $ system, are described. Moreover, a short description of the SFP model is given~\cite{Satchler1979,Etminan:2019gds}.

The S-wave and spin $2$ $\Omega N$ potential in configuration space
	has been obtained by the HAL QCD Collaboration with nearly physical quark masses 
	\cite{Iritani2019prb}. The discrete lattice potential is fitted by an analytic function~\cite{etminan2014}	
	\begin{equation}
	V_{\Omega N}^{HAL}\left(r\right)=b_{1}e^{-b_{2}r^{2}}+b_{3}\left(1-e^{-b_{4}r^{2}}\right)\left(\frac{e^{-m_{\pi}r}}{r}\right)^{2},\label{eq:NOmega_pot}
	\end{equation}	
	where the Gaussian functions describe the short-range and the Yukawa function corresponds to the meson-exchange potential at medium to long-range distances.
	In Ref.~\cite{Iritani2019prb}, the discrete lattice results are fitted reasonably well with
	$\chi^{2}/d.o.f\simeq1$, for four different sets of parameters $ b_{1,2,3} $ and $ b_{4} $, they are given
	in  Table 1 of Ref.~\cite{Iritani2019prb}, 
	where each set of parameters corresponds to the imaginary-time slices $ t/a=11,12,13, 14 $ and $ a=0.0846 $ fm is the lattice spacing.
	The pion mass $ m_{\pi}=146 $ MeV in Eq.~\eqref{eq:NOmega_pot} is taken from the lattice simulation. 	
	For this potential, the scattering length, effective range  and  binding energy  are  
	 $a_{0}^{\Omega N}=5.30$ fm, $r_{0}^{\Omega N}=1.26$ fm and $B^{\Omega N}=1.54\: $ MeV respectively ~\cite{Iritani2019prb}.

As mentioned in the Introduction, the other $\Omega N$ potential model developed by Sekihara, Kamiya and Hyodo~\cite{Sekihara2018} is the meson-exchange (ME) potential, and it is given in configuration space by,
\begin{equation}
 V_{\Omega N}^{ME}\left(r\right)=\frac{1}{4\pi r}\sum_{n=1}^{9}C_{n}\left(\frac{\Lambda^{2}}{\Lambda^{2}-\mu_{n}^{2}}\right)^{2}\left[\exp\left(-\mu_{n}r\right)-\frac{\left(\Lambda^{2}-\mu_{n}^{2}\right)r+2\Lambda}{2\Lambda}\exp\left(-\Lambda r\right)\right],
	\label{eq:pot-EM-omegaN}
\end{equation}
 where the cutoff parameter $ \Lambda=1 $ GeV, $  \mu_{n}=100\:n $ MeV and the parameters $ C_{n} $ are given by the real part of the last column in Table V of Ref.~\cite{Sekihara2018}.

\begin{figure*}[hbt!]
	\centering
	\includegraphics[scale=1.0]{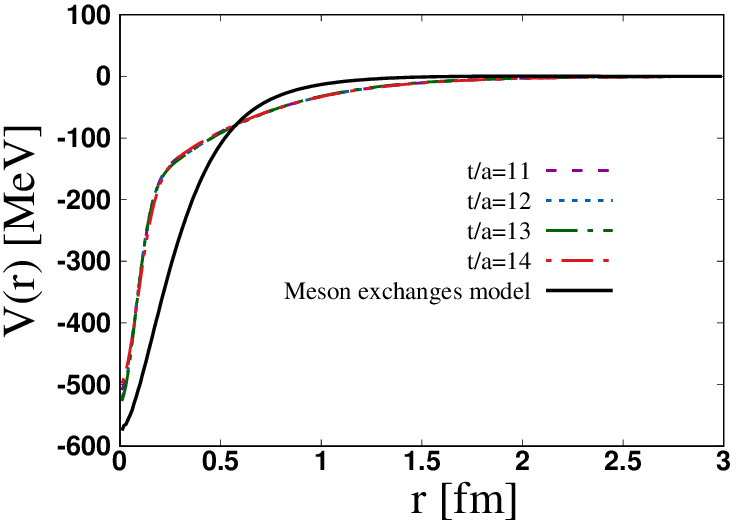}
	\caption{$\Omega N $ potentials as functions of the distance between $ \Omega  $ and $ N $.  HAL QCD $\Omega N $ potential $ V_{\Omega N}^{HAL} $ in Eq.~\eqref{eq:NOmega_pot} at the imaginary-time distances $ t/a=11,12,13,14$ are shown  by parametrization from Ref.~\cite{Iritani2019prb} is compared with the $\Omega N $ potential based on the meson-exchange model (solid black line), $ V_{\Omega N}^{ME} $ in Eq.~\eqref{eq:pot-EM-omegaN} from  Ref.~\cite{Sekihara2018}.
		 It should be mentioned that the HAL potentials are only in the $ ^{5}S_{2} $ channel but in the case of ME potential model the coupled-channel effects are included in the $ \Omega N\:^{5}S_{2} $ interaction.
		  \label{fig:OmegaN} }
\end{figure*}

In Fig.~\ref{fig:OmegaN}, the $\Omega N$ potential from the  
HAL QCD in Eq.~\eqref{eq:NOmega_pot} at the imaginary-time distances $ t/a=11,12,13,14$ ~\cite{Iritani2019prb} and 
 the meson-exchange model ~\cite{Sekihara2018} are depicted. 
 Fig.~\ref{fig:OmegaN} reveals a qualitative difference between these two models.  
 Therefore, it is desirable to find out how these differences can be studied from the  $\Omega\alpha $ two-particle momentum correlation functions.

The effective $ \Omega +\alpha $ nuclear potential is approximated by the single-folding potential (SFP) model
\begin{equation}
	V_{\Omega \alpha}\left(r\right)=\int\rho\left(r^{\prime}\right)V_{\Omega N}\left(\left|\textbf{r}-\textbf{r}^{\prime}\right|\right)d\textbf{r}^{\prime},\label{eq:V_alfaOmega}
\end{equation}		
 where $V_{\Omega N}\left(\left|\textbf{r}-\textbf{r}^{\prime}\right|\right)$ is $\Omega N$ potential between the $\Omega$-particle at $\textbf{r}$
and the nucleon at $\textbf{r}^{\prime}$~\cite{Satchler1979,Miyamoto2018,Etminan:2019gds}; 
moreover, $\rho\left(r^{\prime}\right)$ is the nucleon density function in
 $\alpha$-particle at a distance $\textbf{r}^{\prime}$ from its
center-of-mass which can be taken as~\cite{Akaishi1986},
\begin{equation}
	\rho\left(r^{\prime}\right)=4\left(\frac{4\beta}{3\pi}\right)^{3/2}\exp\left(-\frac{4}{3}\beta r^{\prime2}\right),\label{eq:nucleon-density}
\end{equation}
 $\beta$ is a constant and it is defined by the root-mean-square (rms) radius of $\textrm{\ensuremath{^{4}}He}$, i.e,
 $ \textrm{\ensuremath{r_{r.m.s}}}=\frac{3}{\sqrt{8\beta}}=1.47\:\textrm{fm} $ ~\cite{Akaishi1986}. 
The integration in Eq.~\eqref{eq:V_alfaOmega} is over all space where
 $\rho\left(r^{\prime}\right)$ is defined. 
\section{Two-particle correlation function}\label{sec:Two-particle-CF}
Two-particle correlation function formalism has been explained in detail in various publications such as~\cite{KOONIN197743,Pratt1986,PhysRevC.91.024916,OHNISHI2016294,cho2017exotic}. 
Here, only the essential formulae are provided. The two-particle momentum correlation
function $C_{q}$ is defined by Koonin-Pratt (KP) formula~\cite{OHNISHI2016294}
\begin{equation}
	C\left(q\right)=\int d\boldsymbol{r}S\left(\boldsymbol{r}\right)\left|\Psi^{\left(-\right)}\left(\boldsymbol{r},\boldsymbol{q}\right)\right|^{2},
\end{equation}
where $S\left(r\right)=\exp\left(-\frac{r^{2}}{4R^{2}}\right)/\left(4\pi R^{2}\right)^{3/2}$ is
the single-particle source function that is assumed to be spherical
and static Gaussian  with  source size (source radius) $ R $. 
	The relative Gaussian source function $ S\left(r\right) $ defines the distribution
	of the $\Omega\alpha$ pair production at the relative distance $ r $.	
	If $R_{\Omega}$ and $R_{\alpha}$ are the source sizes of the single
	$\Omega$ and $\alpha$ emissions, respectively, 
	then the effective  radius of the source is given by $ R=\sqrt{\left(R_{\Omega}^{2}+R_{\alpha}^{2}\right)/2}$ ~\cite{cho2017exotic,kamiya2024}. 
The relative wave function $\Psi^{\left(-\right)}$ contains only the S-wave interaction
effect. 
The resulting correlation function can be written as
\begin{equation}
	C\left(q\right)=1+\int_{0}^{\infty}4\pi r^{2}\:dr\:S\left(r\right)\left[\left|\psi\left(q,r\right)\right|^{2}-\left|j_{0}\left(qr\right)\right|^{2}\right], \label{eq:kp}
\end{equation}
where $j_{l=0}\left(qr\right)=\sin\left(qr\right)/qr$ is the spherical
Bessel function and $\psi\left(k,r\right)$ is the S- wave scattering
wave function. For a given two-body $\Omega\alpha$ potential it can
be obtained straightforwardly by solving the Schr\"{o}dinger equation.

When the source size is much larger than the interaction range,
 it is possible to employ the asymptotic behaviour of the wave function,
$\psi\left(q,r\right)\rightarrow j_{0}\left(qr\right)+f\left(q\right)\exp\left(iqr\right)/r$,
which leads to a much simpler formula for the correlation function
in terms of the scattering length and the effective range, which is usually called
the Lednicky-Lyuboshits (LL) approach~\cite{Lednicky:1981su},
\begin{equation}
	C_{LL}\left(q\right)=1+\frac{\left|f\left(q\right)\right|^{2}}{2R^{2}}F_{0}\left(\frac{r_{0}}{R}\right)+\frac{2\textrm{Re}\:f\left(q\right)}{\sqrt{\pi}R}F_{1}\left(2qR\right)-\frac{\textrm{Im}\:f\left(q\right)}{R}F_{2}\left(2qR\right), \label{eq:ll}
\end{equation}
where  $f\left(q\right)\approx1/\left(-1/a_{0}+r_{0}q^{2}/2-iq\right)$ is the scattering amplitude which can be calculated with the effective range expansion (ERE) formula given below in Eq.~\eqref{eq:ERE}.
Further, $F_{1}\left(x\right)=\int_{0}^{x}dt\:e^{t^{2}-x^{2}}/x,\:F_{2}\left(x\right)=\left(1-e^{-x^{2}}\right)/x,$
and the factor $F_{0}\left(x\right)=1-x/\left(2\sqrt{\pi}\right)$
is a correction due to the deviation of the true wave function from the
asymptotic form~\cite{Lednicky:1981su,OHNISHI2016294}.

\section{Numerical Results} \label{sec:result}		
\begin{figure*}[hbt!]
			\centering
			\includegraphics[scale=1.0]{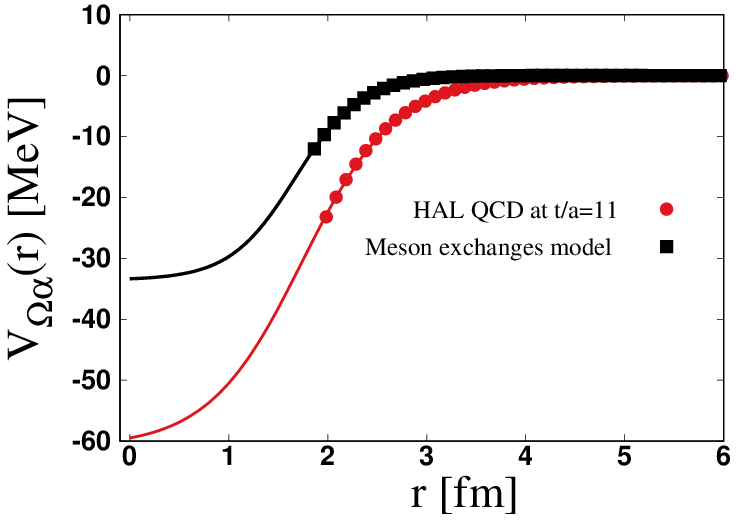}
			\caption{ The single-folding potential $V_{\Omega \alpha}\left(r\right)$ as a function of distance between $ \Omega $ and $\alpha$. Potentials are obtained by evaluating Eq.~\eqref{eq:V_alfaOmega} using $\Omega N $ potential models of HAL QCD at $ t/a=11 $ (red circle) and the meson-exchange model (black square).  
				The corresponding $\Omega N $ potentials are depicted in Fig.~\ref{fig:OmegaN}. 
				In both cases, the solid lines show the fitting of $V_{\Omega \alpha}\left(r\right)$  
				by using $ V_{\Omega \alpha}^{fit}\left(r\right) $ in Eq.~\eqref{eq:ws-fit}. \textcolor{red}{The fit range is taken to be $ r\gtrsim1.9 $ fm~\cite{filikhin2024folding}.}
				The results of the fit and the corresponding parameters are summarized in Table~\ref{tab:Omegaalpha-ERE-SFP}. 		
				  \label{fig:v_OmegaAlpha}}
		\end{figure*}

 \textit{Single-folding $ \Omega \alpha $ potential.}- $\Omega \alpha$ potential is obtained by evaluating Eq.~\eqref{eq:V_alfaOmega} and the resulting potentials are depicted in Fig.~\ref{fig:v_OmegaAlpha} for the HAL QCD (at the imaginary-time distances $ t/a = 11 $) and the meson-exchange  potential models. 
The HAL potential is more attractive  than the ME potential almost at all distances. 
The former is much deeper than the latter and more slowly goes to zero. 
But, in both cases, the interaction ranges are about $ 3 $ fm, although, it is slightly bigger than $ 3 $ fm for the HAL potential.   
			
\begin{figure*}[hbt!]
			\centering
			\includegraphics[scale=1.0]{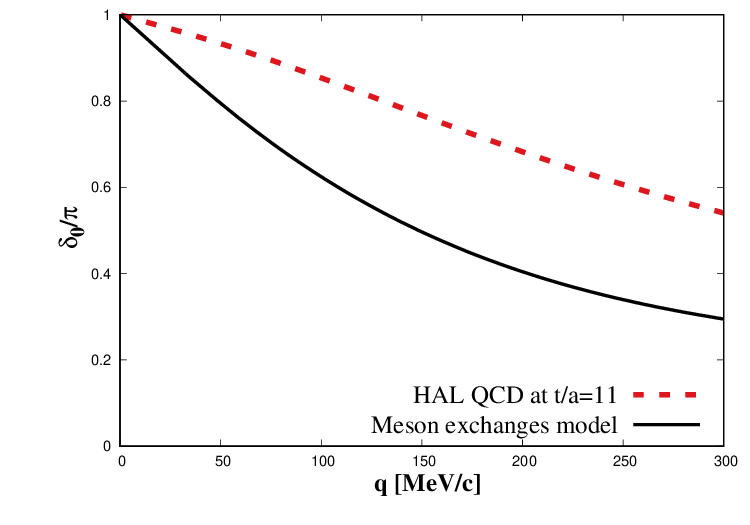}
			\caption{ The normalized $\Omega \alpha$ phase shifts $ \delta_{0} /\pi$  as functions of 
				the relative momentum $ q=\sqrt{2\mu E} $  ($ \mu $ is the reduced mass of $ \Omega\alpha $ system)  using HAL QCD at $ t/a=11 $ (dashed red line) and meson-exchange model (solid black line) potentials.  
				\label{fig:phase_DG_HAL}}
		\end{figure*}

		For phenomenological application and calculation of observables, such as scattering phase shifts and binding energies, I fit $V_{\Omega \alpha}\left(r\right)$ to a Wood-Saxon form using the function that is given by the following equation (motivated by common Dover-Gal model of potential~\cite{dover1983}) with three parameters,  $ V_{0}, R_{c}$ and $c$,			
		\begin{equation}
			V_{\Omega \alpha}^{fit}\left(r\right)=-V_{0}\left[1+\exp\left(\frac{r-R_{c}}{c}\right)\right]^{-1} , \label{eq:ws-fit}
		\end{equation}		
		where $ V_{0}$ is known as the depth parameter,
		\textcolor{red}{ $ R_{c} = r_{c} A^{1/3} $ is the radius of the nucleus		
		 (with the mass number $ A $, i.e, $ A = 4 $ for $ \alpha $) 
		  measured from the center to a point where the density falls to roughly half of its value at the center,		 
		  and the parameter $ c $ is known as the surface diffuseness.  
		  The WS fit for the $\Omega\alpha$ potential is constructed for the interval $r\gtrsim1.9$ fm~\cite{filikhin2024folding}. 
		  This range is chosen according to the rms radii of $^{4}He$ from experimental measurements. 
		  The nucleon density function in the $\alpha$-particle is defined
		  in such a way to reproduce the experimental rms radii. 
		  From electron scattering measurements it is found that,
		  on the average, the radius of a nucleus consisting of $ A $ nucleons
		  is well represented by $R_{Nucl.}=1.2A^{1/3}$.
		  Therefore, the fit interval must be around $ R_{Nucl.}\simeq1.9 $ fm for the $\alpha$-particle.
		  The influence of the rms radius and fitting interval on the $ \Omega \alpha $ potential
		  has recently been thoroughly investigated in~\cite{filikhin2024folding}.	
		  Final fitted parameters $ V_{0}, r_{c} $ and $ c $ are given in Table~\ref{tab:Omegaalpha-ERE-SFP} for the HAL at $ t/a=11 $ and the ME model of potentials.	  
		}
		By using the fit functions (solid lines in Fig~\ref{fig:v_OmegaAlpha}) as input, 
		the Schr\"{o}dinger equation was solved to extract the binding energy and  scattering observables from the asymptotic behaviour of the wave function.
		Figure~\ref{fig:phase_DG_HAL} shows the $\Omega \alpha$ phase shift $ \delta_{0} $ calculated with the potential from the HAL QCD at $ t/a=11 $ and the  meson-exchange potential. 
		The behaviour of the phase shift for both cases shows an attractive interaction to form a bound
		state with a binding energy of about $ 24 $ MeV for the HAL potential and about $ 7.5 $ MeV for the ME potential. 
						
		Low-energy part of $\Omega \alpha$ phase shifts in Fig.~\ref{fig:phase_DG_HAL} provides the scattering length  and the effective range  by employing the ERE formula up to the next-leading-order (NLO),		
		\begin{equation}
			q\cot\delta_{0}=-\frac{1}{a_{0}}+\frac{1}{2}r_{0}q^{2}+\mathcal{O}\left(q^{4}\right).\label{eq:ERE}
		\end{equation}		
			
The fit parameters of scattering length, effective range and binding energy $B_{\Omega \alpha}$, from  HAL and ME potentials are given in Table~\ref{tab:Omegaalpha-ERE-SFP}. The fit functions using Eq.~\eqref{eq:ws-fit} from these parameters are plotted in Fig.~\ref{fig:v_OmegaAlpha} by solid lines. 
As quoted in the caption of Table~\ref{tab:Omegaalpha-ERE-SFP}, the numbers within parentheses correspond to the calculations by using $ \Omega $  mass derived by the lattice simulations~\cite{Iritani2019prb}, where they are slightly larger than the experimental masses.  
Since, by increasing the mass, the contribution of repulsive kinetic energy will decrease and finally lead to slightly deeper binding energies.
 Moreover,  binding energies,  $ B_{\text{\ensuremath{\Omega^{-}}}\alpha^{++}} $ $ \left( B_{\Omega \alpha} \right) $, with (without) Coulomb interaction are given.
  A Coulomb potential due to a uniformly charged sphere is included.  

\begin{table}[hbt!]
\caption{
	The fit parameters of $\Omega \alpha$ potential in Eq.~\eqref{eq:ws-fit}
	and the corresponding low-energy parameter, scattering length $ a_{0} $, effective range $r_{0}$ and binding energy $B_{\Omega \alpha}$, are given for HAL at $ t/a=11 $ and ME model of potentials \textcolor{red}{where the fit range is taken to be $ r\gtrsim1.9 $ fm~\cite{filikhin2024folding}}.
The results have been obtained by using the experimental masses of $\alpha$ and $\Omega$, $3727.38 $ MeV/c and $ 1672.45 $ MeV/c, respectively. 
Furthermore, the results corresponding to $  \Omega $ mass value derived by the HAL QCD Collaboration $ 1711.5 $ MeV/c  are given within parentheses. $ B_{\text{\ensuremath{\Omega^{-}}}\alpha^{++}} $ $\left(B_{\Omega \alpha}\right)$ is the binding energy with (without) Coulomb potential. 
For comparison, the experimental ERE parameters for neutron-neutron scattering are 
$ \left(a_{0},r_{0}\right)=\left(-18.5,2.80\right) $ fm.
\label{tab:Omegaalpha-ERE-SFP}}	
	\begin{tabular}{cccccccc}
		\hline
		\hline 
	Model & $ V_{0} $ (MeV)& $ r_{c} $(fm)& $ c $ (fm)& $a_{0}$(fm)&$r_{0}$(fm)& $B_{\Omega \alpha}$(MeV)& $B_{\text{\ensuremath{\Omega^{-}}}\alpha^{++}}$ (MeV) \\
	\hline
HAL QCD at $ t/a=11 $    & $ 61.0 $& $ 1.10 $ & {$ 0.47 $} & $0.79(0.63)$ & $ 2.81(5.80)$ & $22.9 (23.3)$ & $24.2(24.6)$\\	
Meson-exchange           & $ 33.6 $& $ 1.05 $ & {$ 0.33 $} & $2.65(2.60)$ & $ 1.30(1.28)$ & $ 6.4  (6.6)$ & $ 7.5(7.7)$\\
	\hline
	\hline 	
\end{tabular}
\end{table}
\textit{$ \Omega \alpha $ correlation function.}-
In order to calculate the two-particle correlations from KP formula, Eq.~\eqref{eq:kp}, I used the "\textit{Correlation Analysis tool using the Schr\"{o}dinger Equation}" (CATS)~\cite{cats}. For a given interaction potential and an emission source of any form~\cite{mihaylov2023novel}, CATS is designed to calculate the correlation function.

\begin{figure*}[hbt!]
	\centering
	\includegraphics[scale=0.42]{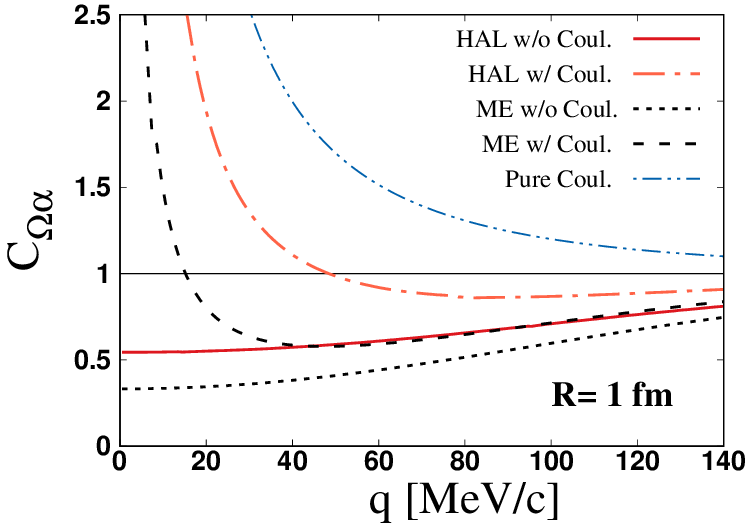} \includegraphics[scale=0.42]{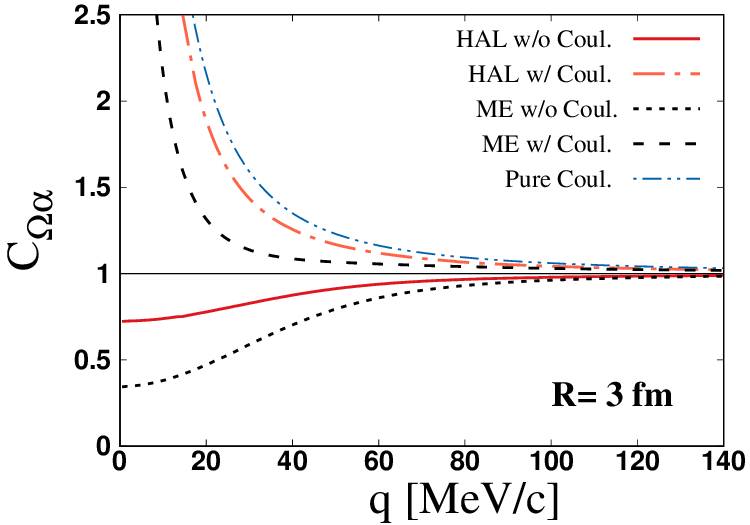} \includegraphics[scale=0.42]{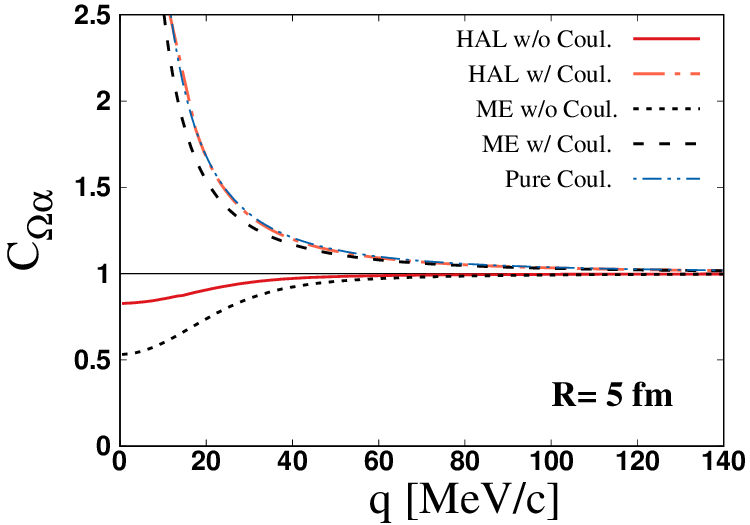}
	\caption{ $ \Omega \alpha $ correlation functions for three different source sizes. The solid and dash-dotted red lines
		are the results from the HAL QCD 
		$ \left(t/a=11\right) $ potentials, without (w/o) and with (w/) Coulomb potential, respectively.  
		The dotted and dashed black lines are the results of the meson-exchange potential, without and with Coulomb potential, respectively.  
		For comparison, the pure Coulomb result, when the strong interaction is switched off, is presented by the dot-dash-dotted blue thin line for comparison.
		\label{fig:omegaAlpha_cq_kp_R}}
\end{figure*}

$ \Omega \alpha $ correlation functions from the two $ \Omega \alpha $ potentials using the KP formula~\eqref{eq:kp} for three different source sizes, $ R=1,3 $ fm and $ 5 $ fm  are calculated and depicted in Fig.~\ref{fig:omegaAlpha_cq_kp_R}, 
where the choice is motivated by values
suggested by analyses of the $ \Lambda \alpha $ correlation function~\cite{jinno2024femtoscopic}.
	Since the charge radius of the $\alpha$-particle is about $1.68 $ fm ~\cite{Krauth2021},
	the source radius of $ R=1 $ fm, may seem small for the emission of the $\alpha$-particle. 	
	But, as discussed in Ref.~\cite{cats}, the term $4\pi r^{2}S\left(r\right)$
	in Eq.~\eqref{eq:kp} describes the probability distribution of the relative distance
	$r$ where the relative source function $S\left(r\right)$ has the
	Gaussian width $\sqrt{2}R$~\cite{cho2017exotic,kamiya2024}. 
	Correspondingly, with the source size $R=1$ fm, the mean distance
	between the emitted pair can be about $\left\langle r\right\rangle =4R/\sqrt{\pi}\sim2.26$ fm that is sufficiently larger than the value of parameter $R$.	
The results with Coulomb attraction are shown by dash-dotted red lines for HAL QCD potential and dashed black lines
for the meson-exchange potential in Fig.~\ref{fig:omegaAlpha_cq_kp_R}.
Once we include the Coulomb interaction between the
negatively charged $ \Omega $ and the positively charged $ \alpha $, a
strong enhancement of $ C(q) $ at small $ q $ is obtained by the long-range attraction. 
For a large source, the distinction of two potentials is smeared by the Coulomb attraction.
The pure Coulomb result, when the strong interaction is switched off, is also illustrated in Fig.~\ref{fig:omegaAlpha_cq_kp_R}.

Also, it can be seen from Fig.~\ref{fig:omegaAlpha_cq_kp_R} that in the area of low momentum $ q\lesssim100 $ MeV/c, the results for two potentials are different.
According to Fig.~\ref{fig:v_OmegaAlpha}, the HAL potential model is more attractive than the ME potential model, thus it gives enhancement of $ C_{\Omega\alpha}\left(q\right) $. But it is rather difficult to get this conclusion from Fig.~\ref{fig:OmegaN}. 
Nevertheless, with the increase of the source size ($ R = 3 $ and $ 5 $ fm), the difference between 
the $ C_{\Omega\alpha}\left(q\right) $s decreases until they are almost the same for $ R=5 $ fm.
Therefore, the future measurement of $ \Omega \alpha $ correlation function from a small source at small relative momentum, can be substantially constrained by $ \Omega N $ interaction at high densities.

\begin{figure*}[hbt!]
	\centering
	\includegraphics[scale=0.42]{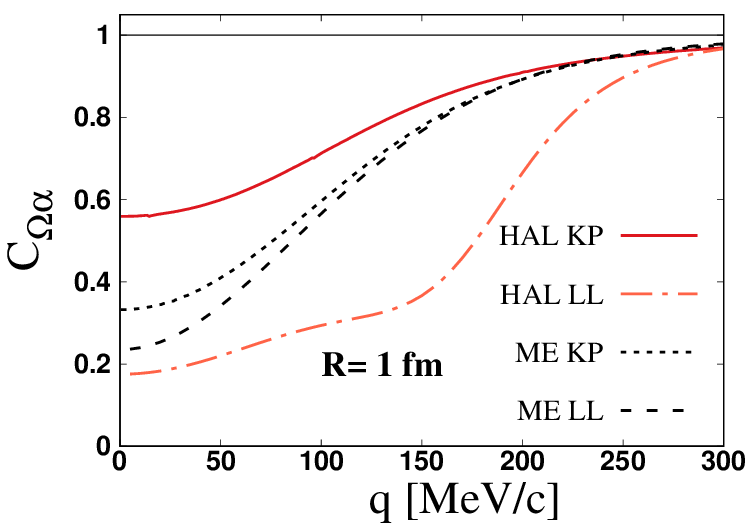} \includegraphics[scale=0.42]{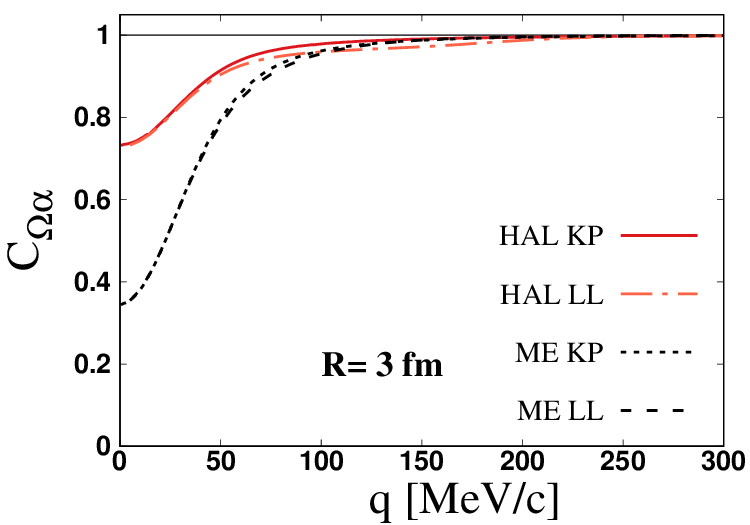} \includegraphics[scale=0.42]{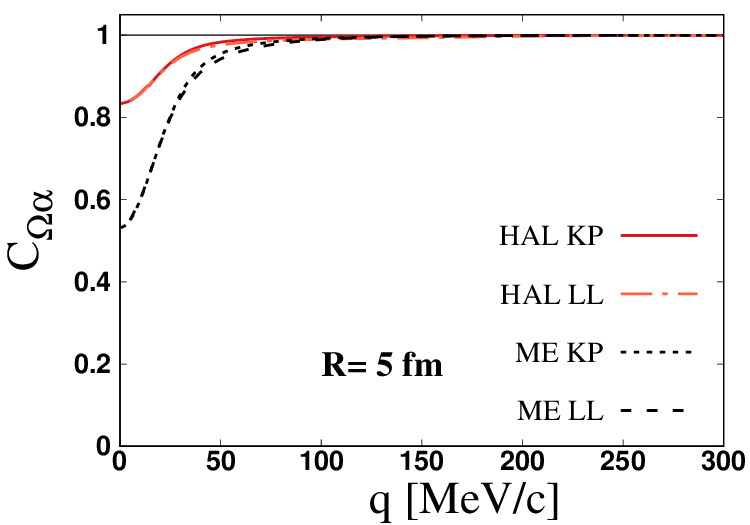}
	\caption{ $ \Omega \alpha $ correlation functions for three different source sizes.
		 The solid and dash-dotted red lines are the results of the HAL QCD 
		$ \left(t/a=11\right) $ potentials with the KP (Eq.~\eqref{eq:kp}) and the LL (Eq.~\eqref{eq:ll}) relations, respectively.  
		The dotted and dashed black lines are the results from the meson-exchange potential with the KP  and the LL formulae, respectively.  
		The Coulomb interaction is switched off in all cases.
		\label{fig:omegaAlpha_cq_kp_ll_R}}
\end{figure*}

By employing the scattering length and the effective range of the two potential models given in Table~\ref{tab:Omegaalpha-ERE-SFP}, the $ \Omega \alpha $  correlation function is calculated 
by using the LL formula (Eq.~\eqref{eq:ll}), and the results are compared with the ones from the KP formula in Fig.~\ref{fig:omegaAlpha_cq_kp_ll_R} for three different source sizes of $ R=1, 3 $ and $ 5 $ fm.  
Figure ~\ref{fig:omegaAlpha_cq_kp_ll_R} demonstrates that for $ R=1 $ fm, the LL formulation produces significant different results compared with the KP formula at low momentum region.
In principle, the LL formula cannot be a good approximation where the source size
 is smaller than the interaction range (for interactions that include nuclei, it can be $ \gtrsim 3  $ fm)~\cite{jinno2024femtoscopic}. 
 On the other hand, the LL approximation is consistent with the KP formula for relatively large source sizes, i.e. $ R\geqq3 $ fm. 
 Note that in Fig.~\ref{fig:omegaAlpha_cq_kp_ll_R}, only the strong interaction is switched on.

\section{Summary and conclusions\label{sec:Summary-and-conclusions}}
In the present paper,  the $ \Omega \alpha $ potential is obtained from two available  $ \Omega N $ interactions, i.e., first principles HAL QCD calculations and the meson-exchange model. For the latter potential, the coupled-channel effect is considered.
$\Omega +\alpha$ potentials were obtained by using the SFP model and then fitted by a Woods-Saxon type function. 
The numerical results showed that $\Omega +\alpha$ potentials based on the HAL potential is much deeper than the one based on the ME potential with binding energies around $ 24 $ and $ 7.5 $ MeV, respectively.  
In both cases, the interaction ranges are about $ 3 $ fm.

I applied femtoscopy technique to predict $ \Omega \alpha $ momentum correlation functions in
 high-energy nuclear collisions to look for an additional and alternative source of knowledge relevant to the
 $ \Omega N $ interaction.  
 Employing two $ \Omega \alpha $ potentials, correlation functions are calculated  using the KP formula for three different source sizes, $ R=1,3 $ fm and $ 5 $ fm. 
  The effect due to different potentials appear in the correlation functions for small source size around $ 1-3 $ fm, with and without considering the Coulomb interaction, while for source size $ R\gtrsim5 $ fm the correlation functions tended to become similar for both $ \Omega \alpha $ potentials with the Coulomb interaction. 
  In all cases without Coulomb potential, the differences still remain significant at short distance. 
  In conclusion, since the correlation functions are sensitive to the behaviour of $ \Omega \alpha $ potential and Coulomb interaction, we could obtain important information about  the interactions of $ \Omega $-particle in dense nuclear medium.
  
Also, 
the binding energy, scattering length and effective range were calculated by solving the
Schr\"{o}dinger equation using the fitted $ \Omega \alpha $ potentials with input from the HAL and the ME potentials. 
Furthermore, correlation functions were examined within the Lednicky-Lyuboshits approach  and compared with the results of using the KP formula when Coulomb interaction is switched off. 
It was seen, as expected, that results in the low-momentum region from the LL formula for small source size ($ 1 $ fm) significantly differ from those from the KP formula.
  
Finally, 
in this theoretical study, 
the selection of source sizes  $ R = 1, 3 $, and $ 5 $ fm is based on previous studies of the two-hadron correlation function in $ pp $ collisions and heavy ion collisions~\cite{jinno2024femtoscopic,kamiya2024}.
The Koonin-Pratt formula, Eq.~\eqref{eq:kp}, is valid while the two correlated particles can be considered as well separated point-like particle. 
In the case of composite particles like $ \alpha $, since there is a possibility of simultaneous formation of alpha particle the effective source size must be larger than those for the emission of any single hadron~\cite{mrowczynski2019hadron,bazak2020production,StanislawPRC2021}.
Therefore, we are basically facing a 5-body problem of two protons, two neutrons and $ \Omega $, and the emergence of alpha particle and its correlation with the $ \Omega $ take place at same time.
 This effect will be considered in future works.	
I hope that these  theoretical studies could help to
design experiments at  FAIR~\cite{cbm}, NICA, and J-PARC HI~\cite{j-park} in future.

\section*{Acknowledgement}
I thank Yuki Kamiya and Asanosuke Jinno
for useful discussions, comments and sharing some data with me.
 I am grateful to the authors and maintainers of "\textit{Correlation Analysis tool using the Schr\"{o}dinger Equation}" (CATS)~\cite{cats}, a modified version of which is used for calculations in this work.	
 Discussions during the long-term workshop, HHIQCD2024 at Yukawa Institute for Theoretical Physics (YITP-T-24-02), were useful as I finished this work.		
\bibliography{Refs.bib}
\end{document}